\documentstyle[aps,prl,twocolumn,epsf]{revtex}
\begin{document}
\draft
\title{Strangelets with finite entropy}
\author{Dan M. Jensen$^{1,2}$ and Jes Madsen$^2$}
\address{$^1$Physics Department, Brookhaven National Laboratory, Upton, NY
11973\\ $^2$Institute of Physics and Astronomy,
University of Aarhus, 
DK-8000 \AA rhus C, Denmark}
\date{July 12, 1995}
\maketitle

\begin{abstract}
Strangelets with non-zero entropy are
studied within the MIT bag model. Explicit account is taken of the
constraints that strangelets must be color neutral and have a fixed
total momentum. In general, masses increase with increasing
entropy per baryon, and the constraints work so as to increase masses
further. This has an important destabilizing effect on strangelets produced in
ultrarelativistic heavy ion collisions.
\end{abstract}


Several ultrarelativistic heavy-ion collision experiments at Brookhaven
and CERN are searching for (meta)stable lumps of roughly equal
numbers of up, down, and strange quarks, so-called strangelets\cite{exp}. If
created, strangelets are characterized by a very low charge-to-mass
ratio, and they could provide one of the best indications of quark-gluon
plasma formation.

An extensive literature\cite{strange}
has studied the properties of strangelets at
zero temperature, but the impact of non-zero entropy (temperature),
which is certainly a condition to be
expected in the hot environment of ultrarelativistic heavy-ion
collisions, has not been investigated in detail.
Clearly, the addition of thermal energy will lead to an increase in
strangelet masses, and this is indeed what is demonstrated below.
Furthermore, additional increases in the energy come about when one
restricts the strangelets to be color singlets, and to have a fixed total
momentum. All of this leads to a destabilization relative to zero entropy
(temperature) calculations, which is of significant importance for the
experimental production and detection of these objects.

In the present investigation we study strangelets in the finite
entropy regime. We use the multiple reflection expansion approach\cite{BB}
within the MIT bag model\cite{degrand}, and in order to study the important
consequences of color singletness and definite momentum in a transparent
manner, we set all quark
masses equal to zero. Since the s-quark mass is expected to be in the
range of 100--300 MeV we thereby get the ``most optimistic'' values
possible (from a production point of view) for strangelet masses etc.
Using zero quark masses and the multiple reflection expansion
allows us to write many of our expressions
in an analytical form, which is more transparent than the numerical
integrals and sums otherwise obtained. On the other hand it prevents us from
showing individual shell-effects in the energy as a function of baryon
number; only the mean effects of the shells are included. Preliminary
results from a finite temperature shell-model calculation by Mustafa and
Ansari\cite{mustafa} (without the color singlet and momentum restrictions)
indicate, that shells are washed out at temperatures exceeding 10--20
MeV, so above this temperature the two approaches should yield identical
results.

For pedagogical reasons we first look at strangelet properties without
imposing restrictions of color singletness and definite momentum. Here
the general expression for the grand potential of particle species $i$ is
\begin{equation}
\Omega_i = \mp g_iT\int_0^\infty dk{{dN}\over{dk}}\ln\left[ 1\pm\exp(-
(\epsilon (k)-\mu )/T)\right]
\end{equation}
where the upper sign is for fermions, the lower for bosons.
$\mu$ and $T$ are the chemical potential and temperature, $k$ is the
particle momentum, $\epsilon$ the corresponding energy, and $g_i$ the
statistical weight. The smoothed
density of states, ${{dN}\over{dk}}$, is given by the multiple
reflection expansion with MIT bag model boundary conditions. For
spherical strangelets characterized by volume $V=4\pi R^3/3$ and
extrinsic curvature $C=8\pi R$ an integration gives per flavor of
massless quarks (including antiquarks)
\begin{equation}
\Omega_q = -\left( {{7\pi^2}\over {60}}T^4+{{\mu^2T^2}\over 2} +
{{\mu^4}\over{4\pi^2}}\right) V
+\left({{T^2}\over{24}}+{{\mu^2}\over{8\pi^2}}\right) C,
\end{equation}
with a corresponding net quark number, i.e. the number of quarks less the
number of anti-quarks,
\begin{equation}
N_q= -\left( {{\partial\Omega_q}\over{\partial\mu}} \right)_{T,V}
= \left( \mu T^2+{{\mu^3}\over{\pi^2}}\right) V
-{{\mu}\over{4\pi^2}}C .
\end{equation}
For gluons
\begin{equation}
\Omega_g = -{{8\pi^2}\over{45}}T^4V + {{4}\over 9}T^2C .
\end{equation}
Here and in the following we often explicitly write
thermodynamical expressions in terms of $\mu$, $T$, $V$, and $C$. One
should notice, that since we concentrate on spherical systems, $C\equiv
8\pi(3/4\pi)^{1/3}V^{1/3}$, so $V$ is the only independent ``shape''
variable. However the use of $C$ makes it more clear where finite-size
corrections enter. Also, $\mu$ and $T$ are sometimes functions of other
variables, such as particle number $N$ and entropy $S$.

The total $\Omega$ can be found from summing the terms above plus the
bag energy $BV$ and other
thermodynamical quantities like the free energy $F$ and the internal energy
$E$, can be derived. For 3 massless quark flavors of equal chemical potential
(this gives the lowest possible energy and electrical neutrality, so
that no Coulomb energy needs to be taken into account)
one finds
\begin{eqnarray}
\Omega(T,V,\mu)&=& \left( -{{19\pi^2}\over{36}}T^4-{3\over 2}\mu^2T^2
-{{3}\over{4\pi^2}}\mu^4 +B\right) V\cr
&&+\left({{41}\over{72}}T^2+{{3}\over{8\pi^2}}\mu^2\right) C,
\label{grand}
\end{eqnarray}
\begin{eqnarray}
F(T,V,N)&=& \left( -{{19\pi^2}\over{36}}T^4+{3\over 2}\mu^2T^2
+{{9}\over{4\pi^2}}\mu^4 +B\right) V\cr
&&+\left({{41}\over{72}}T^2-{{3}\over{8\pi^2}}\mu^2\right) C,
\end{eqnarray}
\begin{eqnarray}
E(S,V,N)&=& \left( {{19\pi^2}\over{12}}T^4+{9\over 2}\mu^2T^2
+{{9}\over{4\pi^2}}\mu^4 +B\right) V\cr
&&-\left({{41}\over{72}}T^2+{{3}\over{8\pi^2}}\mu^2\right) C ,
\label{E}
\end{eqnarray}
where the entropy $S\equiv -\partial\Omega/\partial T|_{V,\mu}$.

Strangelets are in mechanical equilibrium when
$\partial F/\partial V|_{T,N}=\partial \Omega/\partial V|_{T,\mu}=\partial
E/\partial V|_{S,N}=0$, corresponding to
\begin{eqnarray}
BV&=& \left( {{19\pi^2}\over{36}}T^4+{3\over 2}\mu^2T^2
+{{3}\over{4\pi^2}}\mu^4 \right) V\cr
&&-\left({{41}\over{216}}T^2+{{1}\over{8\pi^2}}\mu^2\right) C .
\label{BV}
\end{eqnarray}
Thus in mechanical equilibrium one gets the following expressions for the
grand potential, free energy, internal energy and baryon number:
\begin{equation}
\Omega=\left( {{41}\over{108}}T^2+{{1}\over{4\pi^2}}\mu^2\right) C,
\end{equation}
\begin{equation}
F=\left( 3\mu^2T^2+{{3}\over{\pi^2}}\mu^4\right) V +
\left( {{41}\over{108}}T^2-{{1}\over{2\pi^2}}\mu^2\right) C,
\end{equation}
\begin{equation}
E=4BV,
\label{u4bv}
\end{equation}
\begin{equation}
A=\left( \mu T^2+{{1}\over{\pi^2}}\mu^3\right) V-{\mu\over{4\pi^2}}C.
\end{equation}
Equation (\ref{u4bv}) follows directly from Eqs.\ (\ref{E}) and
(\ref{BV}), and it is in fact a general result for ultrarelativistic
particles in a bag, since the energy density of a relativistic gas is 3
times the particle pressure, which equals $B$, so $E=3BV+BV=4BV$. For
massive quarks this result no longer holds.

Dotted curves in the Figures illustrate the behavior of energy per
baryon as a function of baryon number and temperature or entropy per
baryon derived from the equations above.

\begin{figure}
\epsfbox{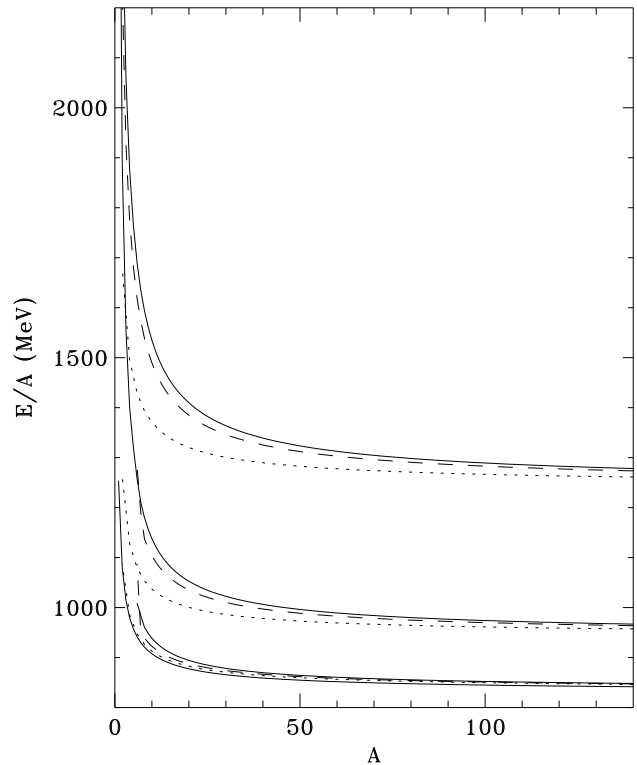}
\caption{
Energy per baryon as a function of baryon number for strangelets with
equal numbers of massless up, down, and strange quarks. $T=0$-results
are shown by the downmost, thin curve. Otherwise, dotted curves are
results without constraints, dashed curves with the color singlet
restriction, and full curves with both color singlet and zero momentum
constraint. The entropy per baryon is 10 for the upper set of curves, 5
for the set in the middle, and 1 for the lowest set. The bag constant
was chosen as $B^{1/4}=145{\rm MeV}$. For other choices of $B$ the
energy scales in proportion to $B^{1/4}$.}
\label{fig1}
\end{figure}

So far we have not explicitly taken into account, that strangelets have
to be color singlets, and must have a definite total momentum. To do this
we use the color singlet and fixed momentum projected grand canonical
partition function of Elze and Greiner\cite{elze}, which we have
independently checked. This partition function, calculated using the
group theoretical projection method\cite{redlich},  is derived in a
saddle-point approximation valid at high temperature and/or chemical
potential (or equivalently baryon density). The partition function is
\begin{equation}
	Z = \Pi_{\text{color}}\Pi_{p=0} Z^{(0)},
\end{equation}
where $\Pi_{\text{color}}$ is the correction factor due to the color
singlet constraint, and $\Pi_{p=0}$ is the correction factor due to
the fixed momentum constraint, here taken at zero total
momentum.  This factorization is only valid in the saddle point
approximation.  $Z^{(0)}$ is the unprojected partition function for a
collection of non-interacting massless quarks, anti-quarks, and gluons in a
spherical MIT bag. (The grand potential in Eq.\ (\ref{grand}) equals
$-T\ln Z^{(0)}$). Both the partition function and the projection
factors are calculated with a density of states based on the multiple
reflection expansion.
The color projection factor is given by
\begin{eqnarray}
	(2\pi\sqrt{3}\, \Pi_{\text{color}})^{-1/4} &=&
	VT^3\left\{2 + {\cal{N}}_q\left[ \frac{1}{3} +\left( \frac{\mu}{\pi T}
	\right) ^2 \right] \right\} \nonumber \\
	&+& CT \frac{12 - {\cal{N}}_q}{12\pi^2},
\end{eqnarray}
and the factor due to the zero-momentum constraint is
\begin{eqnarray}
	\pi\Pi_{p=0}^{-2/3} &=& VT^3\, \pi^2\left\{{\cal{N}}_q\left[
		\frac{7}{30} + \left(\frac{\mu}{\pi T} \right)^2
		+ \frac{1}{2} \left( \frac{\mu}{\pi T} \right)^4
		\right] + \frac{16}{45} \right\} \nonumber \\
		&-& CT\left\{ \frac{{\cal{N}}_q}{72}\left[ 1+3\left(
		\frac{\mu}{ \pi T}\right)^2 \right] +\frac{4}{27} \right\}.
\end{eqnarray}
Terms proportional to ${\cal{N}}_q$, which is the number of massless quark
flavors, originate from quarks, while the remaining terms are due to
gluons.

We now have the ingredients necessary to calculate the energy per
baryon for a zero-momentum, color-neutral
drop of quark matter at finite temperature (entropy). 
As discussed earlier we concentrate on three flavors of massless quarks
with equal chemical potentials.
We introduce the constrained grand potential
\begin{equation}
	\Omega_{\text{con}}(T, V, \mu) = -T \ln Z(T,V,\mu).
\end{equation}
For each baryon number, $A$, we then solve the equations
of mechanical equilibrium
\begin{equation}
	\left( \frac{\partial \Omega_{\text{con}}}{\partial V} 
	\right)_{T, \mu}=0,
\end{equation}
fixed baryon number,
\begin{equation}
	-\left( \frac{\partial \Omega_{\text{con}}}{\partial \mu} 
	\right)_{T,V}=3A,
\end{equation}
and fixed entropy per baryon,
\begin{equation}
	-\frac{1}{A}\left( \frac{\partial \Omega_{\text{con}}}{\partial T} 
	\right)_{V,\mu}=\frac{S}{A},
\label{fixent}
\end{equation}
with respect to $T$, $\mu$, and $V$.

Using $E=4BV$ we then calculate the energy per baryon as
a function of baryon number and show the results for fixed $S/A$ in
Figure 1, where
dashed curves include the color singlet constraint
without the fixed momentum constraint, and full curves include both
color singlet and fixed momentum constraints. As expected (when calculated
for fixed natural variable $S$) both constraints lead to an increase in
energy. For very low $A$ the energy (with constraints included) diverges.
This comes about because the temperature (for fixed $S/A$) increases
above the phase transition temperature for low $A$ and reflects the
break-down of the saddle-point approximation. 

\begin{figure}
\epsfbox{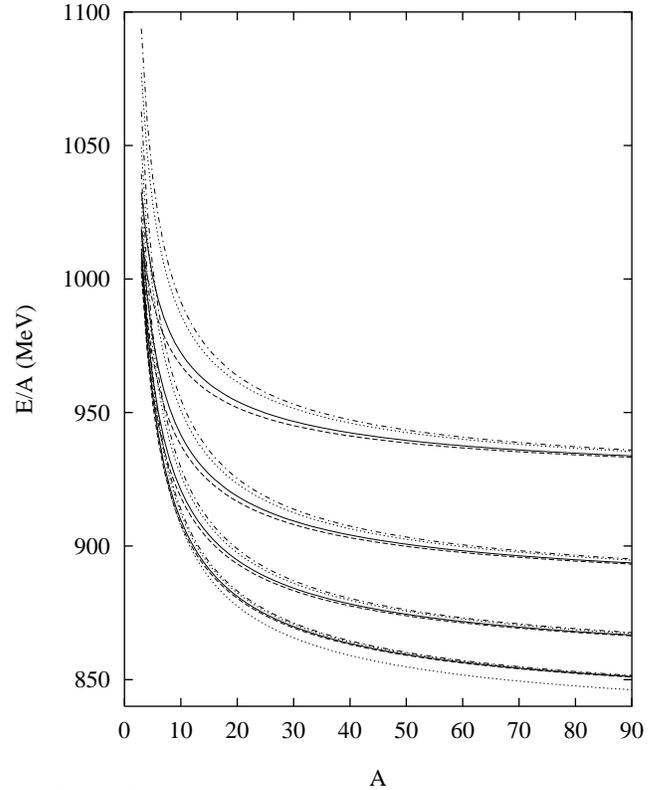}
\caption{The energy per baryon in the unprojected case (dotted lines),
	including the zero-momentum constraint (dashed-dotted lines), including
	the color-singlet constraint (dashed lines), and with both
	constraints (full lines). The calculations were again done for
	$B^{1/4}=145\,\text{MeV}$ and 3 massless quark flavors.
	From bottom to top: $T = 0, 10, 20, 30, 40 \,\text{MeV}$.
	The lower end point of all curves is at $A=3$.}
\end{figure}

Results for fixed $T$ (i.e.\ without imposing Eq.\ (\ref{fixent})) 
are shown in Figure 2, where
the results for the unprojected case, each of the two
projections alone, and together are superimposed for different
temperatures. For sufficiently high
temperature and low baryon number, the effect of (mainly) the color
singlet constraint is equivalent to a lowering of the temperature by
as much as 10 MeV. In other words: the curve including color singlet
corrections (or both corrections) crosses the curves for the
unconstrained calculation at lower temperatures.
It is also seen that the color singlet
constraint is the most important of the two in terms of the effect on
the energy per baryon.

One notices that the effect of color singletness goes away for small
$T$ ($S/A$). This is how it should be, because for $T=0$ there is no
problem in constructing a color neutral strangelet by placing quarks in
the lowest energy levels (e.g. constructing a strangelet with $A=6$ from
2 blue, 2 green, and 2 red up quarks, and similarly for down and strange
quarks, with all quarks in the $1S_{1/2}$ ground state). For $T>0$
quarks are statistically distributed over energy levels, and the
constraints reduce the number of possible configurations, forcing the
energy up. Also, the constraints are only important for $A<100$.

We have shown that the mass of strangelets increase with the entropy per
baryon, or temperature, of the system. 
At fixed entropy per baryon the mass is further
increased when the objects are constrained to be color singlets, and (to
less extent) by the requirement of a definite total momentum (taken to
be zero in the calculations). The total magnitude of the effect is of
order 80 MeV/baryon for temperatures of 40 MeV (which for high baryon
numbers corresponds to roughly 4 units of entropy/baryon), and increases
rapidly for higher entropy (temperature). This important change in
energy (and other corresponding thermodynamical parameters) must be
taken into account in models for production and detection of strangelets
in ultrarelativistic heavy ion collisions. It also plays a role in
relation to quark matter formation in other circumstances, such as
proto-neutron stars.

\acknowledgments
This work was supported in part by the Theoretical Astrophysics Center, under 
the Danish National Research Foundation, and the U.S. Department of Energy 
under contract DE-AC02-76CH00016.

\end{document}